\documentclass[usenatbib]{mnras}

\usepackage{amssymb}
\usepackage{amsmath}
\usepackage{amsfonts}
\usepackage{graphicx}
\usepackage{lipsum}

\newcommand{\mscript}[1]{{\mbox{\scriptsize #1}}}
\newcommand{\aarv}{Ann. Rev. of Astro. and Astrophys.}

\title[A method for evaluating models that use RCs to derive the density profiles]{A method for evaluating models that use galaxy rotation curves to derive the density profiles}

\author[A. F. de Almeida et al]{\'{A}lefe O. F. de Almeida,   Oliver F. Piattella and Davi C. Rodrigues\thanks{E-mail: davi.rodrigues@cosmo-ufes.org} \\
Departamento de F\'{\i}sica, Universidade Federal do Esp\'{\i}rito Santo, Av. Fernando Ferrari 514, Vit\'{o}ria, ES, 29075-910 Brazil}

\begin{document}

\date{}

\pagerange{\pageref{firstpage}--\pageref{lastpage}} 

\maketitle

\label{firstpage}

\begin{abstract}
\noindent
There are some approaches, either based on General Relativity (GR) or modified gravity, that use galaxy rotation curves to derive the  matter density of the corresponding galaxy, and this procedure would either indicate  a partial or a complete elimination  of dark matter in galaxies.  Here we review these approaches, clarify the difficulties on this inverted procedure, present a method for evaluating them, and use it to test  two specific approaches that are based on GR: the Cooperstock-Tieu (CT) and the Balasin-Grumiller (BG) approaches. Using this new method, we find that neither of the tested approaches can satisfactorily fit the observational data without dark matter. The CT approach results can be significantly improved if some dark matter is considered, while for the BG approach no usual dark matter halo can improve its results.
\end{abstract}


\section{Introduction}

Besides the unknown nature of dark matter, the standard model of cosmology ($\Lambda$CDM) is also facing difficulties \citep[e.g.,][]{2009MNRAS.397.1169D, deBlok:2009sp, 2011AJ....142...24O,   BoylanKolchin:2011de, BoylanKolchin:2011dk, Weinberg:2013aya, Pawlowski:2015qta}. There is hope that these issues may be solvable within the $\Lambda$CDM model \citep[e.g.,][]{Governato:2012fa, DelPopolo:2014yta, Onorbe:2015ija}, but the  solutions depend on baryonic physics details with which is  difficult to deal  semi-analytically or through simulations. On the other hand, the answer may be related with more issues than the baryonic physics alone, and may depend on   the nature of dark matter \citep{Moore:1994yx, Colin:2000dn, Hu:2000ke, Zavala:2009ms,Foot:2014uba}, on refinements on the gravitational side \citep{Capozziello:2012ie, Lora:2013fla, Rodrigues:2014xka}, or perhaps on both.

Galaxy rotation curves (RCs) constitute one of the most clear and useful test on the existence of either dark matter or non-Newtonian gravity in galaxies at low redshift.  The determination of the dark matter profile in a galaxy is based on the following schematic procedure \citep[see e.g.,][]{Sofue:2000jx, Courteau:2013cjm}: the observed light is converted into mass densities for the stellar and the gaseous parts. For the stellar part, the conversion depends on the stellar mass-to-light ratio, which depends on the dominant stelar population.  From these mass densities, one derives the corresponding Newtonian potentials, and therefore their individual contributions to the RC. These contributions are typically far from being sufficient to reproduce the observed RC, the difference being attributed to dark matter. In order to compute the dark matter contribution, the usual procedure is to assume a dark matter halo profile that depends on some free parameters which are fitted to the observed RC.

From Newtonian gravity without assumptions on the matter distribution, it is not possible to infer the mass density of a disk galaxy from the observational RC alone, even if it is assumed that all the matter is in a thin axisymmetric disk \citep{0691084459}. What can be done is to compare the RC generated by a given mass density profile, with some free parameters, to the observed RC. 

On the other hand, some non-Newtonian proposals \citep{Cooperstock:2006dt, CoimbraAraujo:2007zz, Dey:2014gka, 2015arXiv150807491M} use the inverse procedure: the observed RC is used as the input from which the mass distribution is derived. For some galaxies, the RC fits of these theories seem satisfactory, but these publications lack a detailed investigation  with respect to the baryonic matter data inferred from observations. Models that use this inverse route have not been yet properly tested and confronted with results from other approaches, it is the purpose of this new method to be able to properly test and compare them. Also, there are other nontrivial approaches that have never been tested with RC data, and this inverse procedure may prove useful to test them \citep[e.g.,][]{Vogt:2007zza,Balasin:2006cg, Rahaman:2008dw,Vieira:2013zba}. 

To address the latter issue, we propose here the effective Newtonian RC method. To exemplify it, two relativistic approaches are selected, the one of the Refs. \citep{Cooperstock:2006dt, Cooperstock:2007sc, Carrick:2011ac}, which we label CT, and the one of Ref.~\citep{Balasin:2006cg}, which we label BG. Both of them use GR in 4D spacetime as the source for gravitational dynamics. Details, merits and criticisms on these approaches are presented in the next sections. This is the first time that the BG approach is studied with realistic galaxy data.

The next section reviews the CT and the BG approaches, sec.~\ref{sec3} presents the effective Newtonian method in generality and its two applications, sec.~\ref{sec4} shows the results, and in sec.~\ref{sec5} we present our conclusions and discussions.

\bigskip

\section{A brief review on two relativistic approaches} \label{sec2}

\subsection{General considerations}

In order to determine the distribution of dark matter in galaxies, the standard approach is to use Newtonian gravity. The motivation for doing so comes usually from the following: $i$) the assumption that GR is the gravitational theory to be considered;  $ii$) that galaxies seem to be stationary systems whose Newtonian potential is small (typically about $\sim 10^{-6}-10^{-8}$, in $c = 1$ units), and $iii$) that the typical speeds are at most about a few hundred km/s (i.e., $\lesssim 10^{-3}$, using $c = 1$ units). These small numbers {\it suggest} that GR corrections to Newtonian dynamics, namely on the rotation curve (RC), are smaller than 1$\%$, and therefore significantly smaller than the typical uncertainties associated with the astrophysical data from galaxies. Some authors agree with the assumption of using GR in galaxies (item $i$), but found that those small numbers in the items $ii$ and $iii$ may lead to significative consequences on dark matter distribution in galaxies and corrections to the RCs larger than 10$\%$ \citep{Cooperstock:2006dt, Carrick:2011ac, 2015arXiv150807491M, Balasin:2006cg, RamosCaro:2012rz}.

The CT approach \citep{Cooperstock:2006ti, Cooperstock:2006dt, Cooperstock:2007sc, Carrick:2011ac, 2015arXiv150807491M} has received a number of criticisms on the theoretical basis, which the authors claim to have answered \citep{Carrick:2011ac}. These criticisms focus on whether the CT approach is indeed fully embedded in GR with a single kind of matter given by a disk of dust. Apart from the theoretical issues, the authors in Refs. \citep{Cooperstock:2006dt, Carrick:2011ac}  also do a strong claim, which is that their approach is capable of reproducing the internal dynamics of about 10 commonly studied galaxies without the need for dark matter, or with only a small amount of it. It appears that, apart from the present work, there is only a single work that criticises the CT phenomenological consequences \citep{Fuchs:2006pr}. Their criticism considers a single galaxy, the Milky Way, and not its rotation curve, but only the velocity dispersion of stars outside the galactic plane or in the Sun neighbourhood. The latter is a valid criticism, but too specific and more prone to observational systematical errors that could erroneously invalidate the model.

The BG approach \citep{Balasin:2006cg} is actually a bifurcation of the CT approach that  avoids certain issues on the CT solution at the galactic plane ($z = 0$) \citep{Vogt:2005va, Balasin:2006cg}. It is claimed that this approach cannot remove dark matter in galaxies, but it can significantly reduce its total amount \citep{Balasin:2006cg}. 

\subsection{The Cooperstock \& Tieu (CT) approach}

The CT approach \citep{Cooperstock:2006dt, Cooperstock:2006ti, Cooperstock:2007sc, Carrick:2011ac, 2015arXiv150807491M} starts from the assumption that all the relevant matter in a  galaxy can be modelled by an axisymmetric stationary dust fluid, and that the spacetime metric can be written as, using $c = 1$ units and standard conventions of the cylindrical coordinates $(r, \phi, z)$,
\begin{equation}\label{metricact}
 ds^2 = -e^{w}(dt-Nd\phi)^2 + e^{-w}r^2d\phi^2 + e^{\nu-w}(dr^2+u \, dz^2), 
\end{equation}  
where $w,N,\nu,u$ are functions that only depend on the coordinates $r,z$. The above line element is not necessarily the most convenient to work on the dynamics of galaxies, but it is the most general with the desired symmetry \citep[e.g., see Chap. 7 of][]{0226870332}.  Since all the considered matter is dust, it is possible to  select coordinates to reduce the number of functions that the metric depends on \citep{Balasin:2006cg}, such that the line element becomes
\begin{equation}\label{metricact}
 ds^2 = -(dt-Nd\phi)^2 + r^2d\phi^2 + e^{\nu}(dr^2 +  dz^2).
\end{equation}  

By either performing a suitable coordinate change \citep{Carrick:2011ac} or an ADM splitting to unveil the lapse function and the shift vector \citep{Balasin:2006cg}, an asymptotic observer at rest with respect to the galaxy center would perceive space rotation with a velocity profile given by
\begin{equation}\label{velocity}
 V= \frac{N}{r}\,.
\end{equation}

The GR field equations impose limits on the form of $N$, in particular from the field equations, 
\begin{equation}\label{eq1first}
N_{rr}+N_{zz}-\frac{N_{r}}{r}=0\:.
\end{equation}
For $z\geq 0$, the following is a valid expression for $V$ \citep{Cooperstock:2006dt},
\begin{equation}\label{velocidadect}
 V_{\mbox{\tiny CT}} (r, z\geq 0)= - \sum_{n}D_{n}e^{-k_{n}  z }J_{1}\left ( k_{n}r \right )\:, 
\end{equation}
where $J_\alpha$ is a Bessel function of the first kind, and $k_n$ and $D_n$ are arbitrary constants. 

The main point of the CT approach is to present a non-Newtonian solution for galaxies that can fit the data without dark matter or a small amount of it, and this non-Newtonian solution needs not to be the most general case. To this end, \citet{Cooperstock:2006dt} have found that they could achieve interesting results by setting the constants $k_n$ to be the $n$-th root of the Bessel function $J_0(k_n \, r_{\mbox{\scriptsize max}})$, where $r_{\mbox{\scriptsize max}}$ is the radius of the farthest observed circular velocity data of a given galaxy. Therefore, at $z=0$, it is not a surprise that eq.~(\ref{velocidadect}) can fit very well the rotation curve of galaxies, it is just a kind of Fourier-Bessel series, which can actually fit almost any curve defined in the interval $(0,r_{\mbox{\scriptsize max}})$. The nontrivial part of eq.~(\ref{velocidadect}) is the $z$ dependence. Hence, it should be clear that the fact that this approach can match very well the observational rotation curve data at $z=0$ is irrelevant for the phenomenology, it is a triviality, since it can fit any curve. What is not trivial is whether the corresponding inferred mass distribution matches the observed baryonic density.

The extension of eq.~(\ref{velocidadect}) by using $ V(r,z) = V(r,-z)$ is a problematic one and was criticised  in particular by \citet{Vogt:2005va, Balasin:2006cg}  \citep[see however][]{Carrick:2011ac}.

The connection between the velocity profile and the matter distribution is derived from the following GR equation,
\begin{equation}\label{eq2}
 \frac{N_{r}^2 + N_{z}^2}{r^2}= 8\pi G \rho \, e^\nu \approx 8\pi G \rho \:.
\end{equation}  

To apply this approach to galaxy RC's, from the RC data one uses eq.~(\ref{velocidadect}) to derive the $D_n$ coefficients. The fit of the curve (\ref{velocidadect}) to the observational RC can be as good as one wants, the higher precision one demands, the larger is the number of $D_n$'s to be fitted. Typically this approach uses about 10 coefficients \citep{Cooperstock:2006dt} (this is just a matter of convention and was found to be suitable to a certain number of galaxies). Hence,  the RC fit alone of this approach is physically irrelevant. The physically important consequences are derived from the matter distribution $\rho$ that is inferred from the RC from eq.~(\ref{eq2}).

The high number of free parameters that the CT approach uses, when confronted to phenomenological results \citep[e.g.,][]{Salucci:2007tm}, seem to indicate alone that either this model is unrealistic or that there must exist some way to reduce its number of free parameters. Indeed, the rule to only pick the  first $n$ roots of the Bessel function is arbitrary. It may exist a rule to select three particular roots of the Bessel function depending on certain galaxy parameters (e.g., the disc scale length), and hence three $D_n$ constants only, which would lead to reasonable galaxy RC fits. Hence, although it is inconvenient that no such rule for selecting the best three $D_n$ parameters is known, the argument on the number of free parameters is not sufficient to discredit this approach.

\subsection{The Balasin \& Grumiller (BG) approach}

The BG approach is a bifurcation of the CT one. It starts from the same line element (\ref{metricact}) and the same energy momentum tensor, but their solution respects the reflection symmetry about the $z=0$ plane, contrary to eq.~(\ref{velocidadect}).

The exact GR field equations derived from the line element (\ref{metricact}) and the energy momentum tensor $T_\mu^\nu = \rho U_\mu U^\nu$ read

\begin{eqnarray}
2r\nu_{r}+ N_{r}^2-N_{z}^2 &=&0\;, \label{eq3} \\
r\nu_{z}+ N_{r}N_{z}&=&0\;, \label{eq4} \\
\nu_{rr}+\nu_{zz}+\frac{1}{2r^2}(N_{r}^2+N_{z}^2)&=&0\;, \label{eq5} \\
N_{rr}+N_{zz}-\frac{N_{r}}{r}&=&0\:, \label{eq1} \\
\frac{N_{r}^2 + N_{z}^2}{r^2}&=& 8\pi G \rho \,  e^{\nu}\:. \label{densitybg}
\end{eqnarray}

\citet{Balasin:2006cg} present the following solution for eq.(\ref{eq1}), 
\begin{equation}\label{solucaobalasin1}
 N(r,z)= A_0 + \int_{0}^{\infty}\cos(\lambda z)(r\lambda)A(\lambda)K_{1}(\lambda r)d\lambda\:,
\end{equation}  
where $A(\lambda)$ is a ``sufficiently regular'' arbitrary function, $A_0$  is a constant and $K_1$ is a modified Bessel function of the second kind. Since the relation between $V$ and $N$ is given by eq.~(\ref{velocity}), it should be  clear that selecting $A(\lambda)$ to fit the observed RC should not be seen as the physical output of this approach, but the physical input. It is shown that a suitable choice of the $A(\lambda)$ function can lead to the following velocity profile,
\begin{eqnarray}\label{velocidadebg}
 V_{\mbox{\tiny BG}}(r,z) =&& \mbox{\hspace*{-0.7cm}}\frac{(R-r_{0})V_{0}}r + \frac{V_{0}}{2r }\sum_{\pm}\left(\sqrt{(z\pm r_{0})^2 + r^2}- \right. \nonumber \\
&&  \mbox{\hspace*{-0.7cm}} \left. - \sqrt{(z\pm R)^2+r^2}\right), 
 \end{eqnarray}
with,  $|z|< r_0$. Hence, at $z=0$,
\begin{equation}\label{velocidadebg}
 V_{\mbox{\tiny BG}}(r,0)= \frac{V_0}r \left ( R-r_{0} + \sqrt{r_{0}^2 + r^2}-\sqrt{ R^2+r^2} \right).
\end{equation}

This profile includes three stages, first the linear increase (for $r \lesssim r_0$), then the constant velocity $V \sim V_0$ regime (for $ r_0 \lesssim r \lesssim R$), and a $1/r$ decrease for $r \gg R$. In practice, for many galaxies, the parameter $R$ cannot be accurately derived from the observational RC, since the transition to a decreasing RC cannot be seen up to the last RC data. 

For circular velocities much lower than the speed of light, and assuming that at $r<r_0$ this approach should coincide with Newtonian gravity, Ref.~\citep{Balasin:2006cg} shows that $\nu$ is (close to) a constant.  Finally, by comparing the differences between Newtonian gravity and their GR approach at the plateau part of the RC, it is argued that this GR approach may significantly reduce the need of dark matter (the estimated differences being about $30 \%$ of the total matter). 

In the following, to simplify the problem in this first step,  we will consider both the CT and the BG approaches without dark matter.
 
\section{The effective Newtonian rotation curve method} \label{sec3}

\subsection{General considerations}

The purpose of this method is to properly and feasibly evaluate models in which the observational RC is used as the model input, while the mass density profile is derived from the latter. The essential feature is to circumvent the use of the commonly unknown matter density error bars by a proper, model dependent, transposition of the observational RC error bars to an effective Newtonian RC. In the end, the method provides an effective RC with error bars that should be fitted with the usual Newtonian procedures.

For the majority of the works on galaxy RC data, and for diverse reasons, there is no profile stating the values of the baryonic density at each radius and its corresponding uncertainty. Therefore, if a model can derive a baryonic density profile by certain means, it is not obvious how to compare it with the expected baryonic profile from the observations. 

The proposed method uses that the relevant uncertainties  are already in part encoded in the RC error bars.   In Refs. \citep[][]{2008AJ....136.2648D, Gentile:2004tb}, like in many others, changes in the redshift data at the same galaxy radius are the main contribution to the RC error bars at that  radius. Therefore, the RC error bars contain information on the violation of perfect axial symmetry, and hence they include information on the maximal confidence one can have on any  axially symmetric model. 

The method that is here proposed depends on the realisation of two minimization procedures. The first one is to derive the model parameters that best fit the observed RC (this fit does not depend on neither the baryonic or the dark matter densities). The second minimization is used to derive the baryonic (and dark matter) parameters, and to yield the relevant quantities to evaluate the goodness of the fit. If one is suspicious on the value of a baryonic model parameter, say the disc scale length, or wants to consider it within a certain range, one can do the last fit with different values of that parameter (without the need of redoing the first fit). The method can be briefly summarised in the following steps: \\
\begin{itemize}

\item The model circular velocity at $z=0$, which is designated by $V(r, p_i)$, where $p_i$ represent the model parameters, is fitted to the observed RC. This RC is described by the table whose $k$-th line reads ($r_k, V_{\mbox{\tiny Obs}, k}, \delta V_{\mbox{\tiny Obs}, k}$), where $r_k$ is the radius of the galaxy whose corresponding circular velocity  is $V_{\mbox{\tiny Obs},k}$ with a 1$\sigma$ error bar given by $\delta V_{\mbox{\tiny Obs},k}$.  The fit determines the  best fit parameters $\bar p_i$ and the corresponding error bars $\delta p_i$. 

\item From $V(r, \bar p_i \pm \delta p_i)$  one can (numerically) determine the corresponding mass density profile as a function of the model parameters, $\rho(r, z, \bar p_i \pm \delta p_i)$, for instance from eqs. (\ref{velocity}, \ref{eq2}).  

\item From $\rho(r, z, \bar p_i \pm \delta p_i)$ one can determine the effective Newtonian circular velocity at $z=0$. The latter is written as $V_{\mbox{\scriptsize eN}}(r, \bar p_i \pm \delta p_i)$ and it is defined as being the circular velocity derived from Newtonian gravity for the matter density $\rho(r,z, \bar p_i \pm \delta p_i)$. 

\item  The effective Newtonian RC data with error bars is built from $V_{\mbox{\scriptsize eN}}$. These data can be expressed as a table  whose $k$-th line is given by ($r_k, \bar V_{\mbox{\scriptsize eN},k}, \delta V_{\mbox{\scriptsize eN}, k}$), where $r_k$ assumes the same values of the original data on the observational RC,  $\bar V_{\mbox{\scriptsize eN},k} = V_{\mbox{\scriptsize eN}}(r_k, \bar p_i)$ and $\delta V_{\mbox{\scriptsize eN},k}$ is an approximation for the  corresponding $1\sigma$ error bar, which is detailed afterwards. 

\item The astrophysical expectation on the gas and stellar densities, together possibly with a given dark matter profile, are used to derive the Newtonian circular velocity $V_{\mbox{\scriptsize N}}$, which will depend on baryonic parameters (like the mass-to-light ratios) and possibly on dark matter parameters as well.

\item If the gravitation theory being considered is compatible with both the observational RC and the matter content assumed for the galaxy, then $V_{\mbox{\scriptsize eN}}$ and $V_{\mbox{\scriptsize N}}$ should be mutually compatible. Hence, one fits $V_{\mbox{\scriptsize N}}$ to the effective Newtonian RC data, thus deriving the baryonic (and dark matter) parameters, and deriving the quantities  $\chi^2$ and $\chi^2_{\mbox{\scriptsize red}}$. The latter are the quantities that have physical information on the quality of the fit and that can be compared to other approaches. 
 
\end{itemize}

\subsection{The fit procedure step by step}\label{subsec:fitprocedure}

Here we describe in detail the procedures associated to the proposed method in four steps:

\vspace{.1in}
\begin{enumerate}
\item {\it The derivation of  $\bar p_i$ and $\rho$.} The minimisation of certain $\chi^2$ is used to compute the best fit parameters for $V(r,p_i)$ in regard to the observed RC data. The corresponding  $\chi^2$ quantity is
\begin{equation}\label{chisquare}
\chi^2_p =\sum_{k=1}^{N}\left(\frac{V(r_{k},p_i)-V_{\mbox{\scriptsize Obs},k}}{\delta V_{\mbox{\scriptsize Obs},k}}\right)^2\;.
\end{equation} 
The subscript $p$ is a reminder that the purpose of $\chi^2_p$ is to derive the model parameters $p_i$ from the observational RC data (this does not constitute the main model results). The values of  $p_i$  that minimise $\chi^2_p$ are denoted by $\bar p_i$, and  $N$ is the total number of  observational data points of the  circular velocity $V_{\mbox{\scriptsize Obs}}$. From the knowledge of $V(r,\bar p_i)$ it is straightforward to evaluate the matter density $\rho(r,z,\bar p_i)$. It is also possible to evaluate $\rho$ for all the values of $p_i$ inside the range given by the error bars  $\delta p_i$, hence one can derive $\rho(r,z,\bar p_i \pm \delta p_i)$. 

\item {\it The derivation of $V_{\mbox{\scriptsize eN}}$.} The effective Newtonian circular velocity can be derived by solving the Poisson equation, $\nabla^ 2\Phi(r,z,\bar p_i)=4\pi G \rho(r,z,\bar p_i)$, and using  $V^{2}_{\mbox{\scriptsize eN}}(r,\bar p_i) =r\partial_r \Phi(r,\bar p_i)$. In particular, the expression for $V_{\mbox{\scriptsize eN}}$ can be directly evaluated from \citep{0691084459}
\begin{eqnarray}
	&& \mbox{\hspace*{-0.7cm}}V^2_{\mbox{\scriptsize eN}}(r,p_i) =  r \, \partial_r \Phi(r,z=0,p_i) \, , \nonumber \\[.1in]
	&& \mbox{\hspace*{-0.7cm}}= -G r \partial_r \int_{-\pi}^\pi d\varphi' \int_{-\infty}^\infty dz' \int_0^\infty dr' \times \nonumber \\
	&&  \;\;\;\;\;\;\;\;\; \times \frac{\rho(r', z', p_i)}{\sqrt{r^2 + {r'}^2  + {z'}^2 - 2 r r' \cos(\varphi')}} \, r'   \, , \nonumber \\[.1in]
	&&  \mbox{\hspace*{-0.7cm}}=  -  2  G  r \int_0^\infty dz' \int_0^\infty dr' \times \label{VeNgeneral} \\
	&& \;\;\;\;\;\;\;\;\; \times \rho(r',z',p_i) \, \partial_r \left( \frac{4 K\left( \frac{4 r r'}{(r + r')^2  + {z'}^2} \right ) }{\sqrt{(r + {r'})^2  + {z'}^2}}\right)  \, r' \, . \nonumber
\end{eqnarray}
In the above, $K$ is the complete elliptic integral defined by
\begin{equation}
	 K(x) = F(\pi/2,x) = \int_0^{\pi/2} d\theta (1 - x \sin^2(\theta))^{-1/2}.
\end{equation}
From the above, it is possible to derive $V_{\mbox{\scriptsize eN}}$ for all the values of $p_i$ inside their 1$\sigma$ uncertainties, that is, one can find $V_{\mbox{\scriptsize eN}}(r, \bar p_i \pm \delta p_i)$. 

\item {\it The effective Newtonian RC.} The purpose of this step is to generate the relevant data, with error bars, that should be used in the next and final fitting procedure. The observational RC is described by the data $(r_k, V_{\mbox{\tiny Obs},k}, \delta V_{\mbox{\tiny Obs},k})$, where $k$ runs from 1 to $N$. The effective Newtonian RC data are given by $(r_k, \bar V_{\mbox{\tiny eN},k}, \delta V_{\mbox{\tiny eN},k})$. In order to avoid the introduction of any bias towards any radii, the same radial values $r_k$ used for the observational RC also appear for the effective Newtonian RC. The quantity $\bar V_{\mbox{\tiny eN},k}$ is simply $ V_{\mbox{\tiny eN}}(r_k,\bar p_i)$ and $\delta V_{\mbox{\tiny eN},k}$ is its corresponding $1\sigma$ error bar. A straightforward procedure to derive the latter goes as follows: firstly one finds $V_{\mbox{\tiny max},k}$ and $V_{\mbox{\tiny min},k}$, which are respectively the maximum and the minimum of $V_{\mbox{\scriptsize eN}}(r_k, p_i)$, with fixed $r_k$,  such that $\chi^2(p_i) \leq \chi^2_{\mbox{\tiny min}} + \Delta \chi^2$, where $\Delta \chi^2$ is the constant associated to a 1$\sigma$ uncertainty considering the total number of the model  parameters ($p_i$). This guarantees that $V_{\mbox{\tiny max},k}$ is the maximum value achievable for $V_{\mbox{\scriptsize eN},k}$ inside the 1$\sigma$ confidence region. Ideally one should  compute the full probability density function (PDF), but depending on the model it may be either exactly valid or be a reasonable  approximation to assume a Gaussian distribution, If the error bars are not exactly symmetric (but are not far from being symmetric), the  $1\sigma$ uncertainty $\delta V_{\mbox{\scriptsize eN},k}$ is set as the maximum between $V_{\mbox{\tiny max},k} - \bar V_{\mbox{\scriptsize eN},k}$ and $\bar V_{\mbox{\scriptsize eN},k} - V_{\mbox{\tiny min},k}$. It should be noted that $\Delta \chi^2$ increases with the number of model parameters $p_i$, and hence in general the larger is the number of parameters $p_i$, the larger will be the uncertainties  $\delta V_{\mbox{\scriptsize eN},k}$.

\item {\it The derivation of the baryonic and dark matter parameters.} Since all the galaxy matter is composed by either baryonic or dark matter, the total Newtonian circular velocity can be expressed by 
\begin{equation}\label{velocitybar}
V^2_{\mbox{\scriptsize N}}=V^2_{\mbox{\scriptsize disk}} + V^2_{\mbox{\scriptsize bulge}}+ V^2_{\mbox{\scriptsize gas}} + V^2_{\mbox{\scriptsize dark matter}}\;.
\end{equation}
To be clear,  $V^2_{\mbox{\scriptsize N}}$  is directly derived from certain matter densities as given above, while $V^2_{\mbox{\scriptsize eN}}$ is derived from the observational RC and from the use of the chosen non-Newtonian gravitation.

For concreteness, here it is considered that the stellar mass-to-light ratios of the bulge and the disk ($\Upsilon_{\mbox{\scriptsize*B}}$ and $\Upsilon_{\mbox{\scriptsize*D}}$) are the only baryonic parameters that are not sufficiently constrained by the observations and need to be fitted. The dark matter contribution will not be considered at the moment, that is, $V^2_{\mbox{\scriptsize dark matter}}=0$. In conclusion, for the assumptions above, $V^2_{\mbox{\scriptsize N}}=V^2_{\mbox{\scriptsize N}} (r, \Upsilon_{\mbox{\scriptsize*B}}, \Upsilon_{\mbox{\scriptsize*D}})$.

If the gravitation theory being considered is compatible with the observational RC and the matter content assumed for the galaxy, then $V_{\mbox{\scriptsize eN}}$ and $V_{\mbox{\scriptsize N}}$ should be mutually compatible. Since $V_{\mbox{\scriptsize N}}$ depends on free parameters, one should evaluate a second and last $\chi^2$ minimization, whose quantity to be minimized reads,
\begin{equation}\label{chisquareeN}
\chi^2 =\sum_{k=1}^{N}\left(\frac{V_{\mbox{\scriptsize N}}(r_{k},\Upsilon_{*\mbox{\scriptsize D}},\Upsilon_{*\mbox{\scriptsize B}})-\bar V_{\mbox{\scriptsize eN},k}}{\delta V_{\mbox{\scriptsize eN},k}}\right)^2\;.
\end{equation}
It is this last $\chi^2$, and the reduced chi-square computed from it ($\chi^2_{\mbox{\scriptsize red}}$),  the quantities that  should be used to compare different approaches, not $\chi^2_p$.

\end{enumerate}

\subsection{Application to the CT approach}

To apply the effective Newtonian method to the CT approach, we follow the steps detailed in  Sec. \ref{subsec:fitprocedure}. \\

\begin{enumerate}
	
\item {\it The derivation of $\bar p_i$ and $\rho$}. The results for $\bar p_i$ and its corresponding error bars can be seen in  Table \ref{tab:parametrospi}. The $p_i$ parameters for this approach correspond to the $D_n$ parameters in eq.~(\ref{velocidadect}). 

For all the six galaxies of this sample we followed the procedure of \citet{Cooperstock:2006dt} of adopting 10 parameters to be fit. For this first fit, the CT approach with 10 parameters could easily fit the observational RC. This can be seen from the values of $\chi^2_{\mbox{\scriptsize p,red}}$ in Table \ref{tab:tabelaresultadoct}. Although 10 parameters is  more than the usual number of parameters used to fit galaxies, this quantity depends on the chosen profile. For the CT approach, which uses eq.~(\ref{velocidadect}), less then five parameters only leads to good fits (i.e.,  $\chi^2_{\mbox{\scriptsize p,red}} \sim 1$) for a few galaxies, typically those whose RC slowly and smoothly increases and hence do not need the high frequency terms of the expansion (\ref{velocidadect}). There are examples in which 10 parameters are not sufficient \citep{2015arXiv150807491M}.

The derivation of $\rho$ from the fitted circular velocity $V$,  at the region with observational RC data, comes from the combination of eqs. (\ref{velocity}, \ref{velocidadect}, \ref{eq2}). 

In general, for the evaluation  of $V_\mscript{eN}$, it is  necessary to consider an extension of $\rho$ beyond the farthest observational RC data, whose radius is $r_{\mscript{max}}$. Namely, the larger is the density beyond $r_\mscript{max}$, the smaller becomes $V_\mscript{eN}$ close to $r_\mscript{max}$ \citep[this is a  known Newtonian effect in axisymmetric systems][]{0691084459}. In principle, one can extend $\rho$ beyond $r_\mscript{max}$ by simply extending the circular velocity curve towards larger $r$ and using eq.~(\ref{velocidadect}). But, as explained in  detail by \citet{Cooperstock:2006ti}, there is no need to use the same $D_n$ and the same $k_n$ beyond $r_\mscript{max}$, and physically reasonable extensions usually require different values for the latter parameters.  From the phenomenological perspective, for sure the baryonic mass density of galaxies must drop at larger radius. As a phenomenologically simple and viable approximation for the total baryonic matter beyond the last observed RC data, we adopt
\begin{equation}
	\rho(r \geq r_{\mbox{\scriptsize max}}, z) = e^{({r_{\mbox{\scriptsize max}} - r})/r_{\mbox{\scriptsize d}}} \rho(r_{\mbox{\scriptsize max}}, z). \label{eq:rhoCTextension}
\end{equation}
This extension is specially natural for the case of a disk galaxy with negligible gas content, since it is just an extension of a Freeman disk \citep{1970ApJ...160..811F}. The gas density usually decays slower than the stellar  component, hence for galaxies with significative amount of gas, the above approximation will cease to be a good one at some radius. Nonetheless, the impact of such deviations on $V_\mscript{eN}$ is  insignificant, since  only the density beyond but close to $r_\mscript{max}$ should contribute significantly to $V_\mscript{eN}$. Moreover, due to the exponential decrease, and the small density at $r_\mscript{max}$, changes on $r_\mscript{d}$ by a factor of two have small or negligible impact on $V_\mscript{eN}$. \\

\item {\it  The derivation of $V_{\mbox{\scriptsize eN}}$}. 
Since, with the extension above, $\rho$ is known in the complete space, deriving $V_{\mbox{\scriptsize eN}}$ reduces to computing the integral (\ref{VeNgeneral}). A technical difficulty can be promptly spotted,  and it comes from the large number of parameters that $V_\mscript{eN}(r,D_n)$ depends on. This difficulty will have consequences to the next step. On the other hand, it is computationally easy to derive the effective Newtonian circular velocity with the best fit $D_n$ parameters, which is written as $\bar V_\mscript{eN}(r)$.\\

\item {\it  The effective Newtonian RC data.}
As detailed in the previous section, these data can be expressed through a table given by $(r_k, \bar V_\mscript{eN,k}, \delta V_\mscript{eN,k} )$, hence at this step one should compute the error bars $\delta V_\mscript{eN,k}$. To this end, it is necessary to perform both a minimization and a maximization of $V_\mscript{eN}(r,D_n)$ with the constraint $\chi^2(p_i) \leq \chi^2_\mscript{min} + \Delta \chi^2$ at each radius $r_k$.  Thus, for each  observational RC data point, and for each one of the six galaxies, one should derive  constrained maximizations and minimizaitons with 10 free parameters.

For the particular case of the CT approach, it is not essencial to compute $\delta V_\mscript{eN,k}$ to conclude that this model (without a large amount of dark matter) cannot describe the astrophysical data of galaxies, since in spite of the error bars values, the $V_\mscript{eN}$ and $V_\mscript{N}$ are systematically incompatible. Moreover, it is not computationally easy to evaluate $\delta V_\mscript{eN,k}$ for the CT approach with ten $D_n$ parameters. 

To evaluate the CT approach with fewer than 10 parameters is helpful as an illustration and to serve as a basis for an estimation of $\delta V\mscript{eN}$ when the 10 parameters are considered.  The galaxy ESO 116-G12 was selected to be analysed with the CT approach and with only three $D_n$ parameters. The results are in Fig. \ref{resultadosct3} and Table \ref{tab:tabelaresultadoct3}. The derived values of $\delta V_\mscript{eN}$ ranges from 0.6 km/s  to 4.5 km/s. With the exception of the first point, all the others have $\delta V_\mscript{eN}$ larger than 1 km/s. The mean $\delta V_\mscript{eN}$ is about 3 km/s.

\begin{figure*}
\begin{minipage}{\textwidth}
\centering
\includegraphics[width=0.9\textwidth]{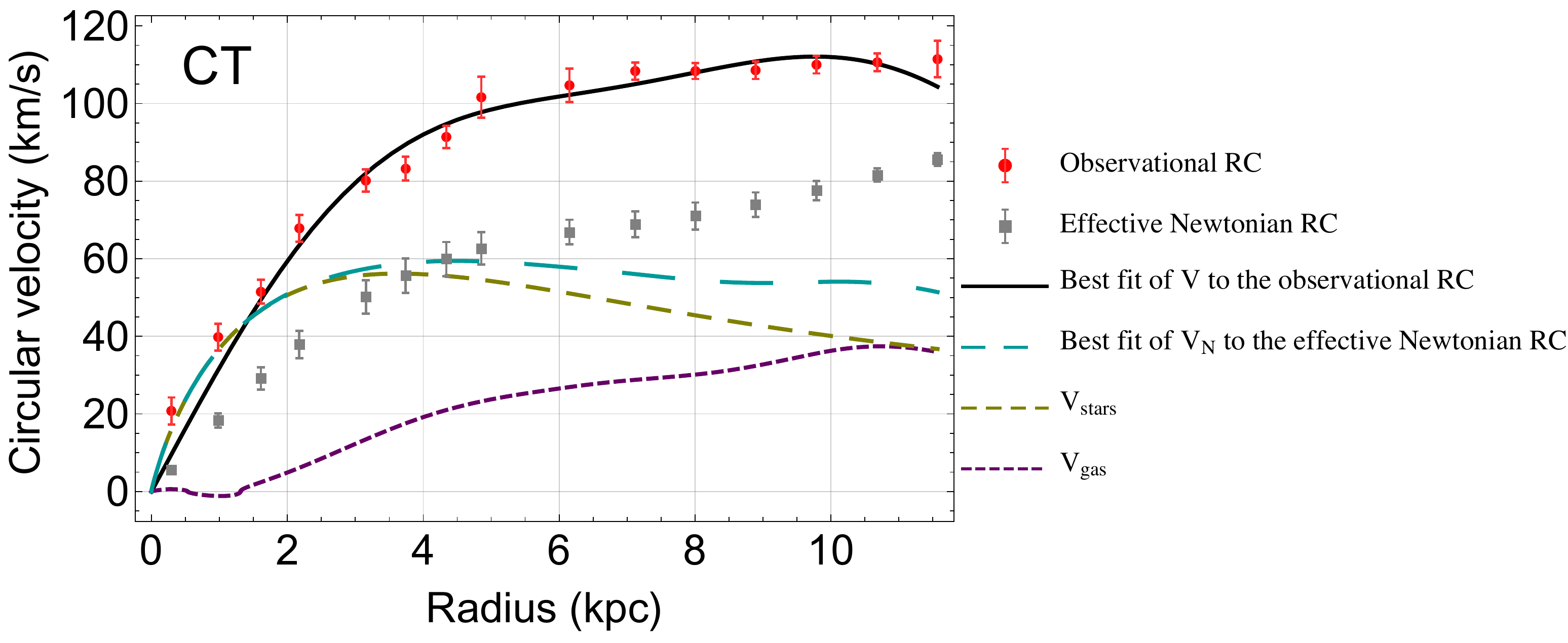}
\caption{\small{The RC curve analysis of ESO 116-G12 with the CT approach with three free model  parameters ($p_i$).  The fit of the total Newtonian circular velocity $V_{\mbox{\scriptsize N}}$ to the effective Newtonian RC is not satisfactory ($\chi^2_{\mbox{ \scriptsize red}} \gg 10 $). }}
\label{resultadosct3}
\end{minipage}
\end{figure*}

\begin{table*}
\begin{minipage}{\textwidth}
\def\arraystretch{1.0}
\setlength\tabcolsep{0.17cm}
\caption{\small Results of the CT approach with three model parameters $p_i$ applied to the galaxy ESO 116-G12 (see Fig. \ref{resultadosct3}). This fit considers the full evaluation of the effective Newtonian data with error bars. This table also includes a comparison to the corresponding results when the same galaxy is modelled with baryonic matter and a NFW dark matter halo  \citep[NFW results from][]{Rodrigues:2014xka}.} 
\label{tab:tabelaresultadoct3}
\centering
\begin{tabular}{lcccccccccc}
\hline
\hline
Galaxy && \multicolumn{4}{c}{CT (3 parameters $p_i$)}  & &  \multicolumn{2}{c}{NFW}    \\
\cline{1-1} \cline{3-6} \cline{8-9} 
   && $\chi^2_{\mbox{\scriptsize{p}}} $    &$\chi^{2}_{\mbox{\scriptsize{p,red}}}$ &$\chi^2$   &$\chi^2_{\mbox{\scriptsize{red}}}$ & & $\chi^2 $& $\chi^2_{\mbox{\scriptsize{red}}}$     \\  
   ESO 116-G12  && 31.60& 2.63& 1341.35& 95.81 & & 31.15& 2.60 &  \\ \hline
  
\end{tabular}
\end{minipage}
\end{table*}

\item {\it  The derivation of the baryonic parameters.} From the fit of $V_\mscript{N}$ to the effective Newtonian RC, one derives $\chi^2$, $\chi^2_\mscript{red}$, $\Upsilon_{* \mscript{B}}$ and $\Upsilon_{*\mscript{D}}$. This is a straightforward procedure, and the results are presented and commented in the next section.
\end{enumerate}

\subsection{Application to the BG approach}

It is easier to apply the effective Newtonian method to the BG than to the CT one for some reasons. The numerical integrals are faster to compute, the model always use 3  parameters $(p_i)$ instead of 10, and the extension of $\rho$ beyond the last observed RC data, $r_\mscript{max}$, is already included as part of this model. \\

\begin{enumerate}
\item {\it  The derivation of $\bar p_i$ and $\rho$}. The results for $\bar p_i$ and its corresponding error bars are displayed in  Table \ref{tab:parametrospi}. The $p_i$ parameters for this approach correspond to the three parameters in eq.~(\ref{velocidadebg}), i.e., $R$, $r_0$ and $V_0$. This first fit that fixes the $p_i$ parameters yields  values for $\chi^2_{\mscript{p,red}}$ in the range from $0.5$ to $1.6$, thus indicating that the velocity profile of this approach is reasonable for describing the RC of galaxies.

The derivation of $\rho$ from the fitted circular velocity $V$ comes from the combination of eqs. (\ref{velocity}, \ref{densitybg}, \ref{velocidadebg}). The extension of  $\rho$ beyond $r_\mscript{max}$ is direct in this BG approach, and essentially it depends on a single parameter ($R$). For some galaxies the value of this parameter can be constrained to lie within some kpc's, but for others, specially those whose RC is monotonously increasing up to $r_\mscript{max}$, there is no maximum for $R$. \\

\item{\it The derivation of $V_\mscript{eN}$.} The effective Newtonian circular velocity $V_\mscript{eN}$ is directly computed from eq. (\ref{VeNgeneral}). \\

\item {\it   The effective Newtonian RC data.} At this step one should compute the error bars $\delta V_\mscript{eN,k}$. To this end, it is necessary to perform both a minimisation and a maximisation of $V_\mscript{eN}(r,r_0, V_0,R)$ with the constraint $\chi^2(r_0,V_0,R) \leq \chi^2_\mscript{min} + \Delta \chi^2$ at each radius $r_k$.  Thus, for each  observational RC data point, and for each one of the six galaxies, one should derive  constrained maximisations and minimisations with 3 free parameters. The derived error bars were either symmetric or close to symmetric, and they were all symmetrized taking the largest value. This was done for all the six galaxies that are in this work evaluated.

\bigskip

\item {\it  The derivation of the baryonic parameters.} From the fit of $V_\mscript{N}$ to the effective Newtonian RC, one derives $\chi^2$, $\chi^2_\mscript{red}$, $\Upsilon_{* \mscript{B}}$ and $\Upsilon_{*\mscript{D}}$. This is a straightforward procedure, and the results are presented and commented in the next section.
\end{enumerate}

\section{Results} \label{sec4}

The results of the fit procedures for CT and BG are in the Tables \ref{tab:parametrospi}, \ref{tab:tabelaresultadoct} and \ref{tab:tabelaresultadobg}, and the RC plots are shown in Fig.~\ref{resultados1}.   

\begin{table*}
\begin{minipage}{\textwidth}	
\caption{The values of the $p_i$ parameters and its errors ($\delta p_i$) for both the CT and BG approaches.}
\label{tab:parametrospi}
\centering
\begin{tabular}{lrrrrrrrrrrrr}
\hline
\hline
\multicolumn{13}{c}{CT approach}\\
\hline
parameters& & DDO 154 & & ESO 116-G12  & & ESO 287-G13 & & NGC 2403 2D& & NGC 2841& & NGC 3198 1D \\
\cline{1-1}\cline{3-3} \cline{5-5} \cline{7-7}\cline{9-9}\cline{11-11}\cline{13-13}  
$D_{1} \left(\mbox{km/s}\right)$& &$303\pm 10$& &$962\pm30$& &$3455\pm 87$& & $1864\pm 13$& &$1.155\pm0.017$ & &$4407\pm 54$\\   
$D_{2}$ (km/s)& &$6.3^{+5.7}_{-5.8}$& &$34\pm22$& &$200^{+61}_{-69}$& & $111.2\pm8.3$& &$1303^{+91}_{-90}$ & &$415\pm31$     \\   
$D_{3}$ (km/s)&  &$10.4\pm4.4$& &$27\pm13$& & $191^{+47}_{-46}$  & &$103.5\pm6.3$& & $928^{+78}_{-76}$& &$280\pm23$ \\  
$D_{4}$ (km/s)&  &$-1.1\pm3.5$& &$-4\pm14$& &$22\pm 44$  & &$26.5\pm5.5$& & $252\pm55$& &$55\pm21$ \\
$D_{5}$ (km/s)&  &$4.8\pm3.0$& &$9.7^{+9.6}_{-9.7}$& &$60^{+49}_{-43}$  & &$13.4\pm5.0$& & $293\pm58$& &$21.9\pm5.9$ \\
$D_{6}$ (km/s)&  &$-2.1\pm2.8$& &$1\pm11$& &$12\pm23$  & &$11.4\pm4.5$& & $101^{+46}_{-47}$& &$5\pm16$ \\ 
$D_{7}$ (km/s)&  &$1.4^{+2.7}_{-2.6}$& &$2.1^{+8.0}_{-7.9}$& &$1\pm43$  & &$7.1\pm4.1$& & $112^{+46}_{-47}$& &$26\pm14$ \\ 
$D_{8}$ (km/s)&  &$-0.3^{+2.3}_{-2.4}$& &$-1.4\pm8.4$& &$3^{+16}_{-15}$& &$7.5\pm3.8$& & $17^{+40}_{-41}$& &$9\pm14$ \\
$D_{9}$ (km/s)&  &$0.7\pm2.0$& &$0.6\pm7.5$& &$16^{+29}_{-28}$& &$8.7\pm3.5$& & $33^{+39}_{-38}$& &$11\pm13$ \\
$D_{10}$ (km/s)&  &$0.2\pm1.5$& &$1.9\pm6.5$& &$-3^{+18}_{-16}$& &$7.9\pm3.2$& & $60^{+28}_{-27}$& &$-0.1\pm11$ \\ [.2in] 
\multicolumn{13}{c}{BG approach}\\
\hline
parameters & & DDO 154 & & ESO 116-G12  & &ESO 287-G13 & & NGC 2403 2D& & NGC 2841& & NGC 3198 1D \\
\cline{1-1}\cline{3-3} \cline{5-5} \cline{7-7}\cline{9-9}\cline{11-11}\cline{13-13}  \\
$R \;\left(\mbox{kpc}\right)$& &$2.1^{+\infty}_{-2.1 }\times 10^7$& & $63^{+\infty}_{-40}$& &$6.7^{+\infty}_{-6.7}\times 10^{7}$& & $4.37^{+\infty}_{-0.10}\times 10^7$& &$109^{+13}_{-11}$ & &$78.9^{+12}_{-9.3}$\\[.2cm]
   $r_0\;\left(\mbox{kpc}\right)$& &$1.18^{+0.13}_{-0.12}$& &$1.79^{+0.56}_{-0.38}$& &$1.308^{+0.095}_{-0.090}$& & $0.706^{+0.034}_{-0.033}$& &$0.31^{+0.14}_{-0.13}$ & &$2.01^{+0.19}_{-0.18}$     \\[.2cm]
      $V_{0}\;\left(\mbox{km/s}\right)$&  &$58.3^{+5.1}_{-1.8}$& &$146^{+40}_{-19}$& & $191.9^{+6.2}_{-2.3}$  & &$141.14^{+0.77}_{-0.76}$& & $345.4^{+7.1}_{-6.8}$& &$197.3^{+6.8}_{-6.3}$     \\[.2cm]
\hline
\end{tabular}
\end{minipage}
\end{table*}

\begin{table*}
\begin{minipage}{\textwidth}	
\def\arraystretch{1.0}
\setlength\tabcolsep{0.12cm}
\caption{\small Results on $\chi^2$ and related quantities of the CT and BG approaches. The corresponding plots are in Fig. \ref{resultados1} \citep[NFW results from][]{Rodrigues:2014xka}.} 
\label{tab:tabelaresultadoct}
\centering
\begin{tabular}{lrrrrrrrrrrrrr}
\hline
\hline
Galaxy && \multicolumn{5}{c}{CT (10 parameters $p_i$)}  & &  \multicolumn{2}{c}{BG}  & & \multicolumn{3}{c}{NFW} \\
\cline{1-1} \cline{3-6} \cline{8-11} \cline{13-14}
  & & $\chi^2_{\mbox{\scriptsize{p}}}$    &$\chi^{2}_{\mbox{\scriptsize{p,red}}}$ &$\chi^2$    &$\chi^2_{\mbox{\scriptsize{red}}}$& &$\chi^2_{\mbox{\scriptsize{p}}}$  &$\chi^{2}_{\mbox{\scriptsize{p,red}}}$ & $\chi^2$& $\chi^2_{\mbox{\scriptsize{red}}}$& &$\chi^2 $  &$\chi^{2}_{\mbox{\scriptsize{red}}}$  \\  
   DDO 154  && 5.93 & 0.12& 103.56& 1.73 & & 26.46 & 0.45& 53.29 & 0.89& & 50.42& 0.87 \\
   ESO 116-G12  && 8.84 & 1.77& 200.75& 14.34& & 63.38 & 1.22& 12.36& 0.88& & 31.15& 2.60 \\
   ESO 287-G13  && 13.96& 0.87& 358.70& 14.35& & 37.68& 1.63& 2278.12 & 91.12& & 36.33& 1.58 \\
   NGC 2403 2D  && 239.29& 0.86& 9200.02& 32.17& & 275.66 & 0.96& 17802.40 & 62.25& & 155.59& 0.55 \\
   NGC 2841  && 58.18& 0.44& 23468.80& 168.84& & 75.79& 0.55& 147.06& 1.06& & 26.52& 0.19 \\
   NGC 3198 1D  && 22.14& 0.26& 2918.39& 31.38& & 59.54& 0.65& 3733.66& 40.14& & 115.67& 1.27 \\    \hline
\end{tabular}
\end{minipage}
\end{table*}

The CT approach with 10 parameters needs at least about $ 10^5$ times more computational time than the BG approach, hence some approximation for $\delta V_\mscript{eN}$ was necessary. All the error bars on this approach were taken to be the same with a common value of  4.5 km/s, based on the maximum error value of the case with 3 parameters presented in Fig. \ref{resultadosct3}. The latter  is plausible since additional parameters on this approach only add Bessel functions of higher frequency, and hence the  error bars derived with ten free parameters are not expected to be much larger then the three parameters ones. 

The plots in Fig. (\ref{resultados1}) clearly show  that, for all the galaxies modelled with the CT approach, the form of the effective Newtonian RC (the grey squares in the plots) systematically does not match the form of the Newtonian circular velocity $V_\mscript{N}$. For all these six galaxies, the best fit Newtonian circular velocity $V_\mscript{N}$ is too high for small radii, and becomes too low at large radii. The latter behaviour is a clear indication that adding a dark matter halo would significantly improve the fit. By adding dark matter, the effective Newtonian RC is the same, but $V_\mscript{N}$ changes by the addition of the new component whose most significative contribution to the RC appears at large radii. 

For the BG approach, there is no  evidence of the same CT systematics. However, there is a less significative tendency with  the opposite behaviour, that is, the $V_\mscript{N}$ curve is  too high at large radii. Hence, no significative improvement on the fits are expected if some   dark matter profile is considered (at least considering the usual dark matter profiles whose density profile decreases much slower than the baryonic density).

The values of $\chi^2_\mscript{p}$ in Table \ref{tab:tabelaresultadoct}  are significantly lower for the CT approach than for the BG one. This is  expected, since the first has more free parameters to fit the observational RC. A reduced $\chi^2$ analysis indicates that the BG approach fits better the observational RC, in the sense that its $\chi^2_\mscript{red,p}$ values are closer to 1.

A good fit related to $\chi^2_\mscript{p}$ is just a minimum requirement for the proposed model to work, it is not sufficient to show that the model is a good one. For the CT and the BG approaches, the $\chi^2$ values are associated to the effective Newtonian RC fit. It is the fit related to $\chi^2$ that is the physically meaningful fit. 

\begin{table*}
\begin{minipage}{\textwidth}
\def\arraystretch{1.0}
\setlength\tabcolsep{0.17cm}
\caption{\small Results on  the stellar mass-to-light ratios of the CT and  BG approaches shown in comparison with  the NFW profile results and the expected values from stellar population considerations. Expected values on $\Upsilon_*$ and NFW results are from Refs.~\citep{Gentile:2004tb, 2008AJ....136.2648D, Rodrigues:2014xka}.} \label{tab:tabelaresultadobg}
\centering
\begin{tabular}{lcccccccccccc}
\hline
\hline
Galaxy && \multicolumn{2}{c}{CT (10 parameters)} & &  \multicolumn{2}{c}{BG}  & & \multicolumn{2}{c}{NFW}& &\multicolumn{2}{c}{Expected} \\
\cline{1-1} \cline{3-4} \cline{6-7} \cline{9-10}\cline{12-13}
   &&$\Upsilon_{\mbox{\scriptsize{*D}}}$&$\Upsilon_{\mbox{\scriptsize{*B}}}$& & $\Upsilon_{\mbox{\scriptsize{*D}}}$&$\Upsilon_{\mbox{\scriptsize{*B}}}$& &$\Upsilon_{\mbox{\scriptsize{*D}}}$&$\Upsilon_{\mbox{\scriptsize{*B}}}$& &$\langle\Upsilon_{\mbox{\scriptsize{*D}}}\rangle$&$\langle\Upsilon_{\mbox{\scriptsize{*B}}}\rangle$\\  
   DDO 154     & & 4.18& -   & & 3.24 & -   & & 1.25&  -  & & 0.2-0.6   &-     \\
   ESO 116-G12  && 0.80& -   & & 0.55 & -   & & 0.05&  -  & & 0.5-1.8&-     \\
   ESO 287-G13  && 1.16& -   & & 0.63 & -   & & 1.69&  -  & & 0.5-1.8&-     \\
   NGC 2403 2D  && 2.39& 0.00& & 0.23 & 2.17& & 0.32& 0.63& & 0.2-0.8   & 0.3-1.2 \\
   NGC 2841     && 1.66& 0.00& & 0.005& 0.24& & 0.72& 1.28& & 0.4-1.5   & 0.4-1.7 \\ 
   NGC 3198 1D    & & 1.07& -& & 0.41& -& & 0.51& -& & 0.4-1.6   & - \\ \hline
\end{tabular}
\end{minipage}	
\end{table*}

\begin{figure*}
\begin{minipage}{\textwidth}
\centering
\includegraphics[trim={8cm 0 8cm 0}, width=16cm]{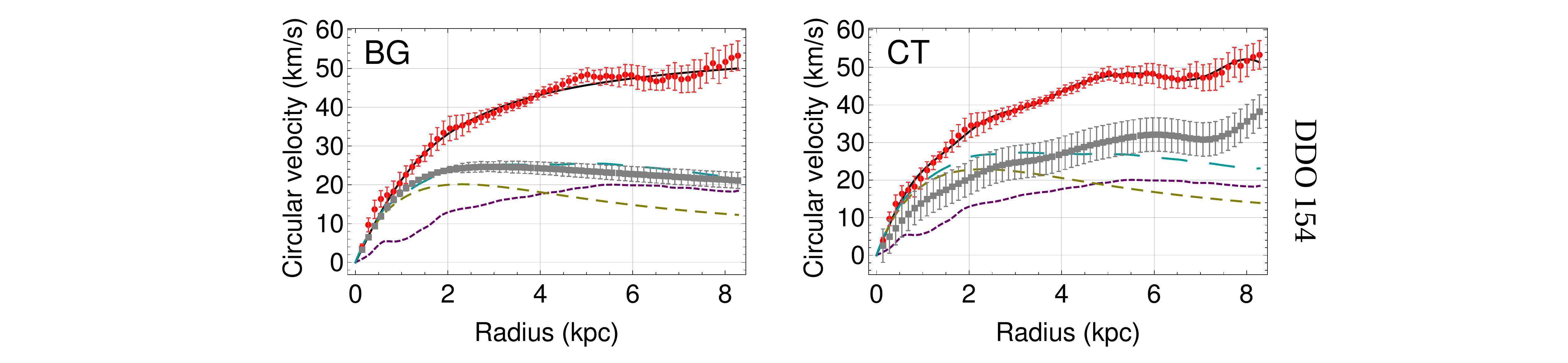}
\includegraphics[trim={8cm 0 8cm 0}, width=15.8cm]{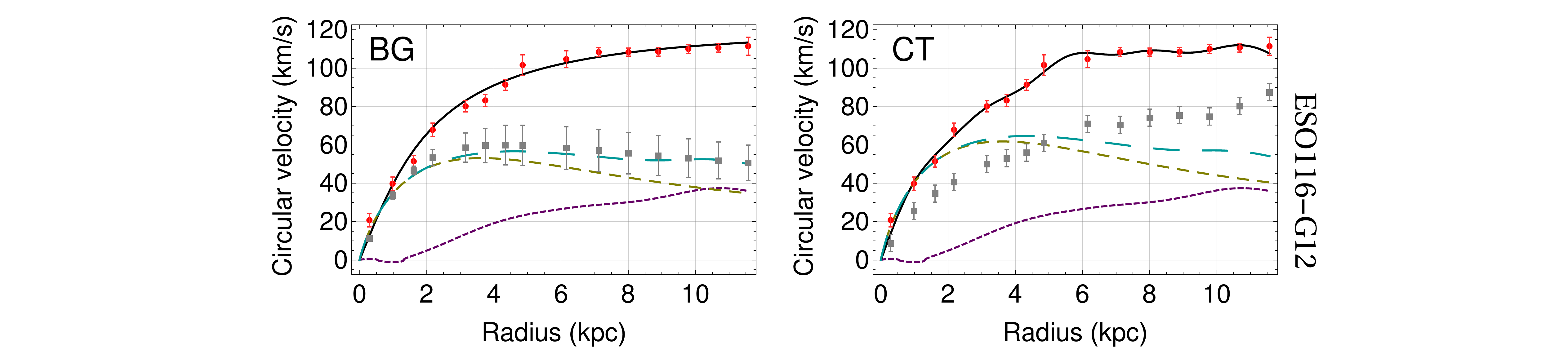}
\includegraphics[trim={8cm 0 8cm 0}, width=15.5cm]{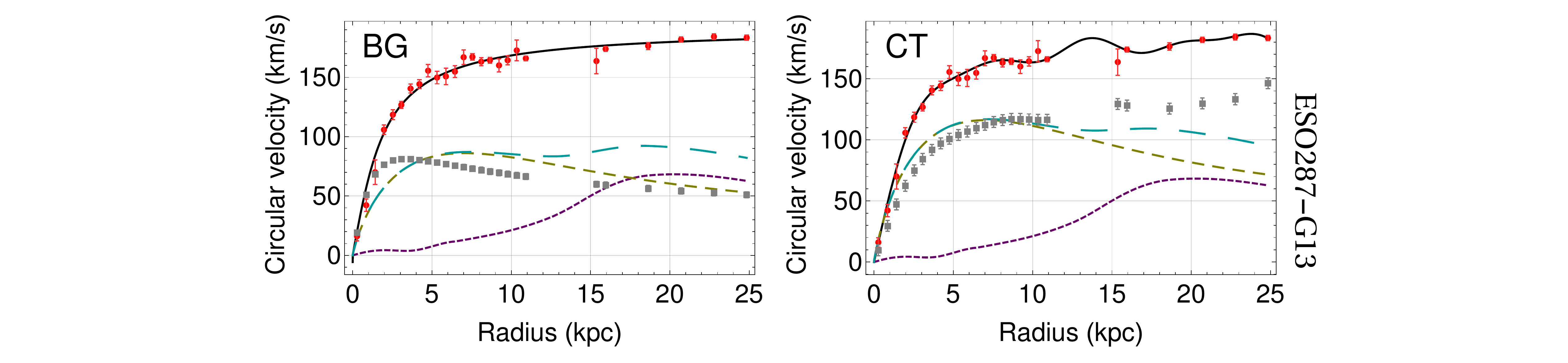}
\includegraphics[trim={8cm 0 8cm 0}, width=15.5cm]{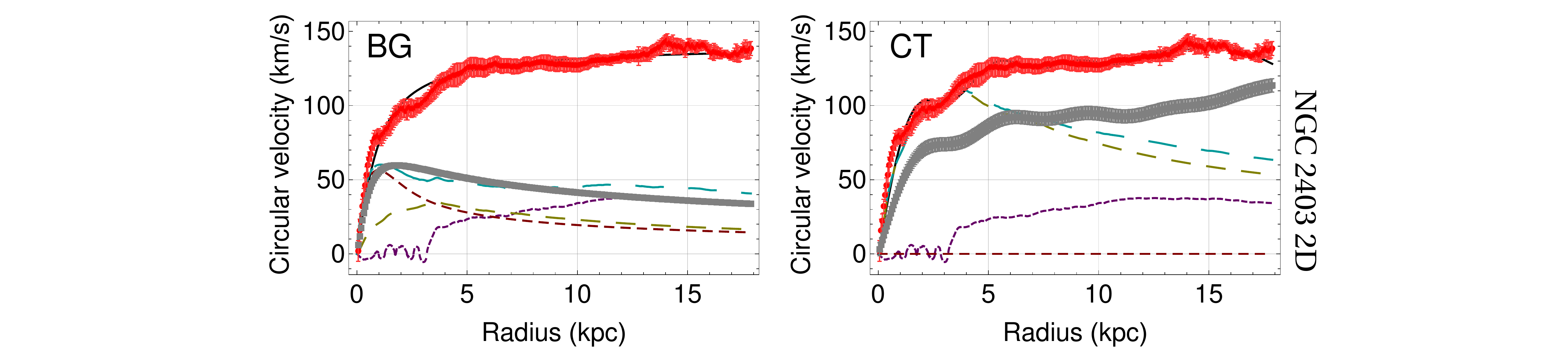}
\caption{\small{Rotation curves  of the galaxies DDO 154, ESO 116-G12, ESO 287-G13 and NGC 2403 2D. The plots use the same conventions of Fig. \ref{resultadosct3}}, with the addition that the dashed dark red curve, when present, refers to the bulge circular velocity ($V_{\mbox{\scriptsize bulge}}$).}
\label{resultados1}
\end{minipage}
\end{figure*}

\begin{figure*}
\begin{minipage}{\textwidth}
\centering
\includegraphics[trim={8cm 0 8cm 0}, width=16cm]{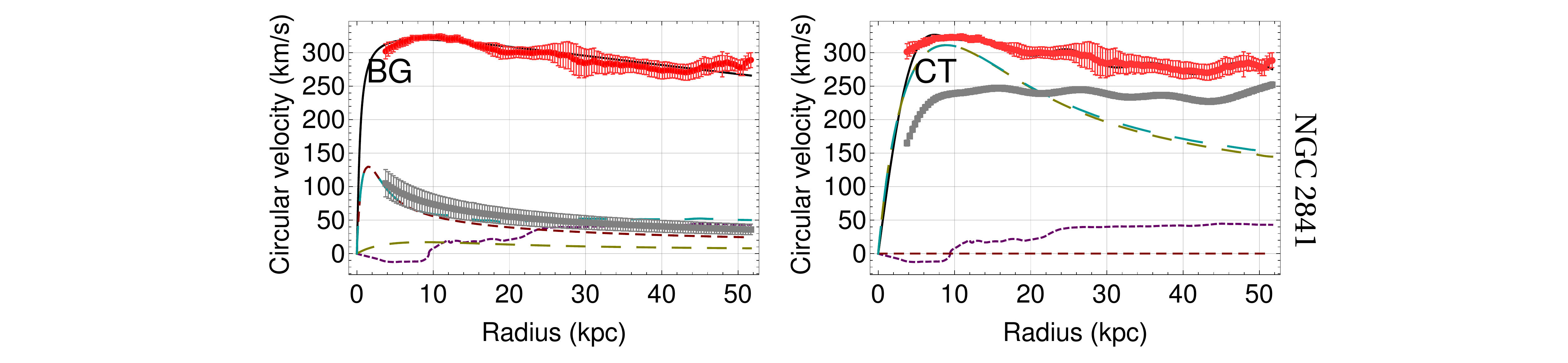}
\includegraphics[trim={8cm 0 8cm 0}, width=15.8cm]{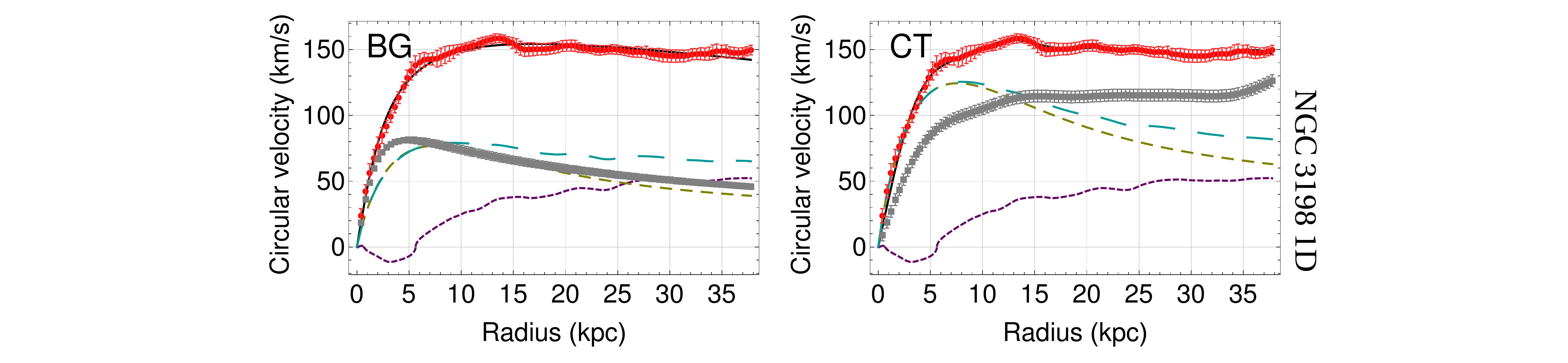}
\contcaption{\small{Rotation curves  of the galaxies NGC 2841 and NGC 3198 1D.}}
\end{minipage}
\end{figure*}

Table \ref{tab:tabelaresultadobg} shows the stellar mass-to-light ratios. The expected ranges for ESO 116-G12 and ESO 287-G13 are the same stated by \citet{Gentile:2004tb}. The other galaxies expectations come from \citet{2008AJ....136.2648D}. We considered a factor two of uncertainty to generate the stated ranges in this table \citep{Bell:2000jt, Meidt:2014mqa}, hence the lower bound is found by dividing the expected value from \citet{2008AJ....136.2648D} by two, and the upper bound by multiplying it by two. The CT approach has a  tendency towards higher $\Upsilon_*$ values, while the BG one tends towards low $\Upsilon_*$ values. This indicates that, by adding a dark matter halo to these approaches, the CT one may benefit from it, achieving better agreement with the expected  $\Upsilon_*$ values, but the BG approach cannot improve and may worsen the  $\Upsilon_*$  concordance if the presence of dark matter is considered.

\section{Conclusions and Discussion} \label{sec5}

There are some models that use the observational rotation curve (RC) data of galaxies to derive the matter density, while the usual route is the opposite one. This class of models have appeared in the context of pure General Relativity (GR) in four dimensional spacetime \citep[e.g.][]{Cooperstock:2006dt, Balasin:2006cg},  in higher dimensions \citep[e.g.,][]{CoimbraAraujo:2007zz}, or in GR extensions \citep[e.g.,][]{Dey:2014gka}. Not always the corresponding papers have clearly stated that they were doing this inversion, and sometimes a simple and successful fit to the observational RC was claimed as an evidence that the proposal can model the internal dynamics of galaxies, even without dark matter. To simply mimic the form of some galaxy RCs is not sufficient (this is one of the general criticisms of \citet{Salucci:2005ip}). 

Here we propose a new method to properly evaluate the application of these approaches when confronted to observational galaxy data. The method only relies on data that can be commonly found in publications on galaxy RCs. Namely, it depends on the observational RC data, the stellar density profile  and the gaseous density profile. 

The method consists of converting the observational RC into a dataset of an effective RC that should be fitted using standard Newtonian gravity procedures. The latter dataset is called the effective Newtonian RC. This conversion is model dependent. The method can be used to consider non-Newtonian gravity models with or without some dark matter. 

The method was here applied to two approaches based on GR, the CT \citep{Cooperstock:2006dt} and the BG \citep{Balasin:2006cg} approaches. Merits and issues related to the theoretical basis of these approaches were briefly reviewed in sec. \ref{sec2}. Our focus was here on testing their phenomenological results, in particular since some publications related to the CT approach \citep[e.g.,][]{Cooperstock:2006dt, Carrick:2011ac}  stated that this approach is capable of achieving good results on the RC fitting of diverse galaxies without dark matter (or with a small amount of it). The BG approach was here confronted with the astrophysical data of galaxies for the first time.  

 The application of the effective Newtonian RC method has shown that both the approaches have strong problems fitting galaxy RCs without dark matter (the selected sample  favours the BG approach over the CT one). The method also indicates that if dark matter is considered, the BG approach cannot improve its results significantly, but the CT approach  can. 

For an example,  we consider the case of the BG approach applied to the galaxy ESO 287-G13, Fig.~\ref{resultados1}. The BG RC could match nicely the observational RC, but is effective Newtonian RC at large radii is smaller than the contribution from the gas alone, hence this model and the data are not compatible (this can also be seen from its large $\chi^2_\mscript{red}$ value at Table \ref{tab:tabelaresultadoct}). Moreover, by adding dark matter to the analyses, the problem is not alleviated, but it increases. By adding any dark matter halo whose most significative RC contribution is at large radii, the Newtonian curve $V_\mscript{N}$ (the dashed cyan curve in the plots) will  become higher at large radii. This is the opposite to what happens for the same galaxy with the CT approach. For the latter, the $V_\mscript{N}$ curve is below the effective Newtonian RC at large radii.

Beyond the two GR approaches tested here, we expect that the evaluation of other models  could benefit from the method that was here introduced.

\section*{Acknowledgements}
We thank Fabio Iocco for critically reading and commenting on a previous version of this paper;  W.J.G. de Blok and G. Gentile for providing relevant data on galaxies that are used in this paper; Fred Cooperstock and Steven Tieu for clarifying an issue we had with their galaxy fitting procedure and  Alberto Saa for relevant discussions on the subject. AOFA thanks CAPES (Brazil) for support. OFP and DCR thank CNPq (Brazil) and FAPES (Brazil) for partial support.

\bibliographystyle{mnras} 

\bibliography{bibdavi2014B}{}

\begin{thebibliography}{}
\makeatletter
\relax
\def\mn@urlcharsother{\let\do\@makeother \do\$\do\&\do\#\do\^\do\_\do\%\do\~}
\def\mn@doi{\begingroup\mn@urlcharsother \@ifnextchar [ {\mn@doi@}
  {\mn@doi@[]}}
\def\mn@doi@[#1]#2{\def\@tempa{#1}\ifx\@tempa\@empty \href
  {http://dx.doi.org/#2} {doi:#2}\else \href {http://dx.doi.org/#2} {#1}\fi
  \endgroup}
\def\mn@eprint#1#2{\mn@eprint@#1:#2::\@nil}
\def\mn@eprint@arXiv#1{\href {http://arxiv.org/abs/#1} {{\tt arXiv:#1}}}
\def\mn@eprint@dblp#1{\href {http://dblp.uni-trier.de/rec/bibtex/#1.xml}
  {dblp:#1}}
\def\mn@eprint@#1:#2:#3:#4\@nil{\def\@tempa {#1}\def\@tempb {#2}\def\@tempc
  {#3}\ifx \@tempc \@empty \let \@tempc \@tempb \let \@tempb \@tempa \fi \ifx
  \@tempb \@empty \def\@tempb {arXiv}\fi \@ifundefined
  {mn@eprint@\@tempb}{\@tempb:\@tempc}{\expandafter \expandafter \csname
  mn@eprint@\@tempb\endcsname \expandafter{\@tempc}}}

\bibitem[\protect\citeauthoryear{Balasin \& Grumiller}{Balasin \&
  Grumiller}{2008}]{Balasin:2006cg}
Balasin H.,  Grumiller D.,  2008, Int.J.Mod.Phys., D17, 475

\bibitem[\protect\citeauthoryear{Bell \& de Jong}{Bell \&
  de~Jong}{2001}]{Bell:2000jt}
Bell E.~F.,  de Jong R.~S.,  2001, \mn@doi [\apj] {10.1086/319728}, 550, 212

\bibitem[\protect\citeauthoryear{Binney \& Tremaine}{Binney \&
  Tremaine}{1988}]{0691084459}
Binney J.,  Tremaine S.,  1988, {Galactic Dynamics (Princeton Series in
  Astrophysics)}.
Princeton University Press

\bibitem[\protect\citeauthoryear{Boylan-Kolchin, Bullock  \&
  Kaplinghat}{Boylan-Kolchin et~al.}{2011}]{BoylanKolchin:2011de}
Boylan-Kolchin M.,  Bullock J.~S.,   Kaplinghat M.,  2011, \mn@doi [\mnras]
  {10.1111/j.1745-3933.2011.01074.x}, 415, L40

\bibitem[\protect\citeauthoryear{Boylan-Kolchin, Bullock  \&
  Kaplinghat}{Boylan-Kolchin et~al.}{2012}]{BoylanKolchin:2011dk}
Boylan-Kolchin M.,  Bullock J.~S.,   Kaplinghat M.,  2012, \mn@doi [\mnras]
  {10.1111/j.1365-2966.2012.20695.x}, 422, 1203

\bibitem[\protect\citeauthoryear{Capozziello \& De~Laurentis}{Capozziello \&
  De~Laurentis}{2012}]{Capozziello:2012ie}
Capozziello S.,  De~Laurentis M.,  2012, \mn@doi [Annalen Phys.]
  {10.1002/andp.201200109}, 524, 545

\bibitem[\protect\citeauthoryear{Carrick \& Cooperstock}{Carrick \&
  Cooperstock}{2012}]{Carrick:2011ac}
Carrick J.,  Cooperstock F.,  2012, Astrophys.Space Sci., 337, 321

\bibitem[\protect\citeauthoryear{Coimbra-Araujo \& Letelier}{Coimbra-Araujo \&
  Letelier}{2007}]{CoimbraAraujo:2007zz}
Coimbra-Araujo C.~H.,  Letelier P.~S.,  2007, \mn@doi [Phys. Rev.]
  {10.1103/PhysRevD.76.043522}, D76, 043522

\bibitem[\protect\citeauthoryear{Colin, Avila-Reese  \& Valenzuela}{Colin
  et~al.}{2000}]{Colin:2000dn}
Colin P.,  Avila-Reese V.,   Valenzuela O.,  2000, \mn@doi [Astrophys. J.]
  {10.1086/317057}, 542, 622

\bibitem[\protect\citeauthoryear{Cooperstock \& Tieu}{Cooperstock \&
  Tieu}{2006}]{Cooperstock:2006ti}
Cooperstock F.,  Tieu S.,  2006, \mn@doi [Mod.Phys.Lett.]
  {10.1142/S0217732306021505}, A21, 2133

\bibitem[\protect\citeauthoryear{Cooperstock \& Tieu}{Cooperstock \&
  Tieu}{2007}]{Cooperstock:2006dt}
Cooperstock F.~I.,  Tieu S.,  2007, Int. J. Mod. Phys., A22, 2293

\bibitem[\protect\citeauthoryear{Cooperstock \& Tieu}{Cooperstock \&
  Tieu}{2008}]{Cooperstock:2007sc}
Cooperstock F.~I.,  Tieu S.,  2008, Mod. Phys. Lett., A23, 1745

\bibitem[\protect\citeauthoryear{Courteau et~al.}{Courteau
  et~al.}{2014}]{Courteau:2013cjm}
Courteau S.,  et~al., 2014, \mn@doi [Rev. Mod. Phys.]
  {10.1103/RevModPhys.86.47}, 86, 47

\bibitem[\protect\citeauthoryear{{Del Popolo}, Lima, Fabris  \& Rodrigues}{{Del
  Popolo} et~al.}{2014}]{DelPopolo:2014yta}
{Del Popolo} A.,  Lima J.,  Fabris J.~C.,   Rodrigues D.~C.,  2014, \mn@doi
  [JCAP] {10.1088/1475-7516/2014/04/021}, 1404, 021

\bibitem[\protect\citeauthoryear{Dey, Bhattacharya  \& Sarkar}{Dey
  et~al.}{2015}]{Dey:2014gka}
Dey D.,  Bhattacharya K.,   Sarkar T.,  2015, \mn@doi [Gen. Rel. Grav.]
  {10.1007/s10714-015-1945-x}, 47, 103

\bibitem[\protect\citeauthoryear{{Donato} et~al.,}{{Donato}
  et~al.}{2009}]{2009MNRAS.397.1169D}
{Donato} F.,  et~al., 2009, \mnras, \href
  {http://adsabs.harvard.edu/abs/2009MNRAS.397.1169D} {397, 1169}

\bibitem[\protect\citeauthoryear{Foot \& Vagnozzi}{Foot \&
  Vagnozzi}{2015}]{Foot:2014uba}
Foot R.,  Vagnozzi S.,  2015, \mn@doi [Phys. Rev.]
  {10.1103/PhysRevD.91.023512}, D91, 023512

\bibitem[\protect\citeauthoryear{{Freeman}}{{Freeman}}{1970}]{1970ApJ...160..811F}
{Freeman} K.~C.,  1970, \mn@doi [\apj] {10.1086/150474}, \href
  {http://adsabs.harvard.edu/abs/1970ApJ...160..811F} {160, 811}

\bibitem[\protect\citeauthoryear{Fuchs \& Phleps}{Fuchs \&
  Phleps}{2006}]{Fuchs:2006pr}
Fuchs B.,  Phleps S.,  2006, \mn@doi [New Astron.]
  {10.1016/j.newast.2006.04.002}, 11, 608

\bibitem[\protect\citeauthoryear{Gentile, Salucci, Klein, Vergani  \&
  Kalberla}{Gentile et~al.}{2004}]{Gentile:2004tb}
Gentile G.,  Salucci P.,  Klein U.,  Vergani D.,   Kalberla P.,  2004, \mnras,
  351, 903

\bibitem[\protect\citeauthoryear{Governato, Zolotov, Pontzen, Christensen, Oh
  et~al.}{Governato et~al.}{2012}]{Governato:2012fa}
Governato F.,  Zolotov A.,  Pontzen A.,  Christensen C.,  Oh S.,   et~al.,
  2012, \mn@doi [\mnras] {10.1111/j.1365-2966.2012.20696.x}, 422, 1231

\bibitem[\protect\citeauthoryear{Hu, Barkana  \& Gruzinov}{Hu
  et~al.}{2000}]{Hu:2000ke}
Hu W.,  Barkana R.,   Gruzinov A.,  2000, \mn@doi [Phys. Rev. Lett.]
  {10.1103/PhysRevLett.85.1158}, 85, 1158

\bibitem[\protect\citeauthoryear{Lora, Grebel, Sanchez-Salcedo  \& Just}{Lora
  et~al.}{2013}]{Lora:2013fla}
Lora V.,  Grebel E.~K.,  Sanchez-Salcedo F.~J.,   Just A.,  2013, \mn@doi
  [Astrophys. J.] {10.1088/0004-637X/777/1/65}, 777, 65

\bibitem[\protect\citeauthoryear{{Magalhaes} \& {Cooperstock}}{{Magalhaes} \&
  {Cooperstock}}{2015}]{2015arXiv150807491M}
{Magalhaes} N.~S.,  {Cooperstock} F.~I.,  2015, preprint, \href
  {http://adsabs.harvard.edu/abs/2015arXiv150807491M} {} (\mn@eprint {arXiv}
  {1508.07491})

\bibitem[\protect\citeauthoryear{Meidt et~al.}{Meidt
  et~al.}{2014}]{Meidt:2014mqa}
Meidt S.~E.,  et~al., 2014, \mn@doi [Astrophys. J.]
  {10.1088/0004-637X/788/2/144}, 788, 144

\bibitem[\protect\citeauthoryear{Moore}{Moore}{1994}]{Moore:1994yx}
Moore B.,  1994, \mn@doi [Nature] {10.1038/370629a0}, 370, 629

\bibitem[\protect\citeauthoryear{{Oh}, {Brook}, {Governato}, {Brinks}, {Mayer},
  {de Blok}, {Brooks}  \& {Walter}}{{Oh} et~al.}{2011}]{2011AJ....142...24O}
{Oh} S.-H.,  {Brook} C.,  {Governato} F.,  {Brinks} E.,  {Mayer} L.,  {de Blok}
  W.~J.~G.,  {Brooks} A.,   {Walter} F.,  2011, \mn@doi [\aj]
  {10.1088/0004-6256/142/1/24}, \href
  {http://adsabs.harvard.edu/abs/2011AJ....142...24O} {142, 24}

\bibitem[\protect\citeauthoryear{O{\~n}orbe, Boylan-Kolchin, Bullock, Hopkins,
  Ker{\v e}s, Faucher-Gigu{\`e}re, Quataert  \& Murray}{O{\~n}orbe
  et~al.}{2015}]{Onorbe:2015ija}
O{\~n}orbe J.,  Boylan-Kolchin M.,  Bullock J.~S.,  Hopkins P.~F.,  Ker{\v e}s
  D.,  Faucher-Gigu{\`e}re C.-A.,  Quataert E.,   Murray N.,  2015, \mn@doi
  [Mon. Not. Roy. Astron. Soc.] {10.1093/mnras/stv2072}, 454, 2092

\bibitem[\protect\citeauthoryear{Pawlowski, Famaey, Merritt  \&
  Kroupa}{Pawlowski et~al.}{2015}]{Pawlowski:2015qta}
Pawlowski M.~S.,  Famaey B.,  Merritt D.,   Kroupa P.,  2015, \mn@doi
  [Astrophys. J.] {10.1088/0004-637X/815/1/19}, 815, 19

\bibitem[\protect\citeauthoryear{Rahaman, Kalam, DeBenedictis, Usmani  \&
  Ray}{Rahaman et~al.}{2008}]{Rahaman:2008dw}
Rahaman F.,  Kalam M.,  DeBenedictis A.,  Usmani A.~A.,   Ray S.,  2008,
  \mn@doi [Mon. Not. Roy. Astron. Soc.] {10.1111/j.1365-2966.2008.13559.x},
  389, 27

\bibitem[\protect\citeauthoryear{Ramos-Caro, Agon  \& Pedraza}{Ramos-Caro
  et~al.}{2012}]{RamosCaro:2012rz}
Ramos-Caro J.,  Agon C.,   Pedraza J.,  2012, \mn@doi [Phys.Rev.]
  {10.1103/PhysRevD.86.043008}, D86, 043008

\bibitem[\protect\citeauthoryear{Rodrigues, de Oliveira, Fabris  \&
  Gentile}{Rodrigues et~al.}{2014}]{Rodrigues:2014xka}
Rodrigues D.~C.,  de Oliveira P.~L.,  Fabris J.~C.,   Gentile G.,  2014,
  \mn@doi [\mnras] {10.1093/mnras/stu2017}, 445, 3823

\bibitem[\protect\citeauthoryear{Salucci \& Gentile}{Salucci \&
  Gentile}{2006}]{Salucci:2005ip}
Salucci P.,  Gentile G.,  2006, \mn@doi [Phys. Rev.]
  {10.1103/PhysRevD.73.128501}, D73, 128501

\bibitem[\protect\citeauthoryear{Salucci, Lapi, Tonini, Gentile, Yegorova  \&
  Klein}{Salucci et~al.}{2007}]{Salucci:2007tm}
Salucci P.,  Lapi A.,  Tonini C.,  Gentile G.,  Yegorova I.,   Klein U.,  2007,
  \mn@doi [Mon. Not. Roy. Astron. Soc.] {10.1111/j.1365-2966.2007.11696.x},
  378, 41

\bibitem[\protect\citeauthoryear{Sofue \& Rubin}{Sofue \&
  Rubin}{2001}]{Sofue:2000jx}
Sofue Y.,  Rubin V.,  2001, \mn@doi [\aarv] {10.1146/annurev.astro.39.1.137},
  39, 137

\bibitem[\protect\citeauthoryear{Vieira \& Letelier}{Vieira \&
  Letelier}{2014}]{Vieira:2013zba}
Vieira R. S.~S.,  Letelier P.~S.,  2014, \mn@doi [Gen. Rel. Grav.]
  {10.1007/s10714-013-1641-7}, 46, 1641

\bibitem[\protect\citeauthoryear{Vogt \& Letelier}{Vogt \&
  Letelier}{2005}]{Vogt:2005va}
Vogt D.,  Letelier P.~S.,  2005, preprint (\mn@eprint {arXiv}
  {astro-ph/0510750})

\bibitem[\protect\citeauthoryear{Vogt \& Letelier}{Vogt \&
  Letelier}{2007}]{Vogt:2007zza}
Vogt D.,  Letelier P.~S.,  2007, \mn@doi [Phys. Rev.]
  {10.1103/PhysRevD.76.084010}, D76, 084010

\bibitem[\protect\citeauthoryear{Wald}{Wald}{1984}]{0226870332}
Wald R.~M.,  1984, {General Relativity}.
University Of Chicago Press

\bibitem[\protect\citeauthoryear{Weinberg, Bullock, Governato, de Naray  \&
  Peter}{Weinberg et~al.}{2013}]{Weinberg:2013aya}
Weinberg D.~H.,  Bullock J.~S.,  Governato F.,  de Naray R.~K.,   Peter A.
  H.~G.,  2013, in {Sackler Colloquium: Dark Matter Universe: On the Threshhold
  of Discovery Irvine, USA, October 18-20, 2012}.  (\mn@eprint {arXiv}
  {1306.0913}), \url
  {http://inspirehep.net/record/1237028/files/arXiv:1306.0913.pdf}

\bibitem[\protect\citeauthoryear{Zavala, Jing, Faltenbacher, Yepes, Hoffman,
  Gottlober  \& Catinella}{Zavala et~al.}{2009}]{Zavala:2009ms}
Zavala J.,  Jing Y.~P.,  Faltenbacher A.,  Yepes G.,  Hoffman Y.,  Gottlober
  S.,   Catinella B.,  2009, \mn@doi [Astrophys. J.]
  {10.1088/0004-637X/700/2/1779}, 700, 1779

\bibitem[\protect\citeauthoryear{de Blok}{de~Blok}{2010}]{deBlok:2009sp}
de Blok W.,  2010, \mn@doi [Adv.Astron.] {10.1155/2010/789293}, 2010, 789293

\bibitem[\protect\citeauthoryear{{de Blok}, {Walter}, {Brinks}, {Trachternach},
  {Oh}  \& {Kennicutt}}{{de Blok} et~al.}{2008}]{2008AJ....136.2648D}
{de Blok} W.~J.~G.,  {Walter} F.,  {Brinks} E.,  {Trachternach} C.,  {Oh} S.,
  {Kennicutt} R.~C.,  2008, \aj, \href
  {http://adsabs.harvard.edu/abs/2008AJ....136.2648D} {136, 2648}

\makeatother
\end{thebibliography}

\bsp
\label{lastpage}
\end{document}